\documentclass[aps,prd,floatfix,superscriptaddress,notitlepage,onecolumn]{revtex4}

\usepackage{epsfig}
\usepackage{amsmath}
\usepackage{graphicx,psfrag}
\usepackage{dcolumn}
\usepackage{bm}
\usepackage{slashed}
\usepackage{xcolor}

\newcommand{\beq}{\begin{equation}}
\newcommand{\eeq}{\end{equation}}
\newcommand{\bea}{\begin{eqnarray}}
\newcommand{\eea}{\end{eqnarray}}

\newcommand\fig[1]     {Fig.\,{\ref{#1}}}

\newcommand\app[1]     {~\ref{#1}}

\def\eq#1{(\ref{#1})}
\def\s0#1#2{\mbox{\small{$ \frac{#1}{#2} $}}}
\def\0#1#2{\frac{#1}{#2}}

\def\mr#1{{\mathrm{#1}}}

\begin{document}
\title{Time Evolution and Thermal Renormalization Group Flow in Cosmology}

\author{Istv\'an G\'abor M\'ari\'an}
\affiliation{HUN-REN Atomki, P.O.Box 51, H-4001 Debrecen, Hungary} 
\affiliation{University of Debrecen, Institute of Physics, P.O.Box 105, H-4010 Debrecen, Hungary}

\author{Andrea Trombettoni}
\affiliation{Department of Physics, University of Trieste, Strada Costiera 11, I-34151 Trieste, Italy}
\affiliation{CNR-IOM DEMOCRITOS Simulation Center, Via Bonomea 265, I-34136 Trieste, Italy}

\author{Istv\'an N\'andori}
\affiliation{University of Debrecen, Institute of Physics, P.O.Box 105, H-4010 Debrecen, Hungary}
\affiliation{HUN-REN Atomki, P.O.Box 51, H-4001 Debrecen, Hungary} 
\affiliation{University of Miskolc, Institute of Physics and Electrical Engineering, H-3515, Miskolc, Hungary}

\date{\today}

\begin{abstract}
  Time-evolution of the Universe as described by the Friedmann equation 
  can be coupled to equations of motion of matter fields. Quantum effects 
  may be  incorporated to improve these classical equations of motion by
  the renormalization group (RG) running of their couplings. Since temporal 
  and thermal evolutions are linked to each other, astrophysical and cosmological
  treatments based on  zero-temperature RG methods require the extension 
  to finite-temperatures. We propose and explore a modification of the usual 
  finite-temperature RG approach by relating the temperature parameter to the 
  running RG scale as $T \equiv k_T = \tau k$ (in natural units), where $k_T$ 
  is acting as a running cutoff for thermal fluctuations and the momentum $k$
  can be used for the quantum fluctuations. In this approach, the temperature 
  of the expanding Universe is related to the dimensionless quantity $\tau$ (and
  not to $k_T$). We show that by this choice dimensionless RG flow equations 
  have no explicit $k$-dependence, as it is convenient. We also discuss how 
  this modified thermal RG is used to handle high-energy divergences of the
  RG running of the cosmological constant and to "solve the triviality"of the 
  $\phi^4$ model by a thermal phase transition in terms of $\tau$ 
  in $d=4$ Euclidean dimensions.
\end{abstract}

%\pacs{98.80.-k, 11.10.Gh, 11.10.Hi}

\maketitle

%===============
\section{Introduction} 
%===============
There is an increasing interest in astrophysics and cosmology to incorporate 
quantum effects by renormalization group (RG) running \cite{wilson} such as in
the asymptotically safe gravity approach \cite{reuter}, for a recent review see 
e.g. \cite{reuter_saueressig,AS_grav_1,AS_grav_2}. Application to cosmology 
\cite{bonanno_saueressig,alessia, rg_cosmo_12, rg_cosmo_17,rg_cosmo_18} 
and to black-hole physics have been extensively discussed \cite{cosmo_qeg,qeg8}. 
In these approaches, the running RG cutoff $k$ is identified with a typical energy 
scale of the system \cite{cosmo_qeg} and the RG running is used to connect the 
physics of different energy scales and at different times.
The RG running of the parameters 
(couplings) is then incorporated into the classical equations of motion,
i.e., the Friedmann 
equations describing the time-evolution of a homogeneous and isotropic universe.

Indeed, particle physics and cosmology are intimately connected, however,
in accelerators a limited number of colliding particles is taken into account, while 
the early Universe must be seen as a hot dense 
plasma. Thus, the zero temperature quantum field theory which is 
the usual theoretical framework for particle processes must be
extended to finite temperatures for cosmological applications.
Thermal field theory is a very well-developed framework for finite
temperature applications \cite{kapusta}. However, 
when applied to cosmological problems one has to decide whether 
setting the temperature is
to be equal to the running RG momentum scale or link it
to a fixed momentum. The latter choice appears to be
the correct one since the quantized theory must be obtained in 
the physical limit where the RG scale is sent to zero, 
nevertheless it has a serious drawback, as we are going to argue in the following. 
One finds an explicit RG scale dependence in 
the dimensionless RG flow equations, and therefore is no room for non-trivial 
fixed points. This is the problem we address in the present paper.

Several types of cutoff identifications \cite{cosmo_qeg} has been 
used in astrophysics and cosmology. By using units in which 
$c = \hbar = k_B = 1$, one can relate the RG scale $k$ to the energy density, 
$k \propto \rho^{1/4}$, or to the temperature of the cosmic plasma, $k \propto T$ 
\cite{alessia}. In the radiation dominated epoch one finds $\rho \propto T^4$,
and thus these are identical to each other. In addition, the RG scale $k$ can be 
connected to the time-dependent Hubble-parameter $H(t)$ which is compatible 
with the Bianchi identities, see e.g.~\cite{alessia}. Then one can introduce the Hubble 
(or expansion) time, $t_{\rm H} = H^{-1} = \sqrt{3/(8 \pi G \rho)} \propto T^{-2}$, 
where $G$ is the Newton constant and we used the relation $\rho \propto T^4$
holding in the radiation dominated epoch. Thus, the temporal and the thermal 
evolution of the Universe are linked to each other, confirming 
that astrophysical and cosmological applications of the RG approach may require 
the extension to finite-temperatures.

In this paper we decided to explore the finite 
temperature generalization of the RG equations written in Euclidean 
spacetime (say, in $d$ dimensions) via the following modification 
for the integrals entering the action:
$\int d^dx \to \int_0^\beta d\tilde t \int d^{d-1}x$, 
where $\beta = 1/T$ is the inverse temperature parameter and
one sees that for $T=0$ the full integration on $d$ variables,
including the imaginary time $\tilde t$, is restored.
It is employed in many-body physics \cite{negele,kapusta} and
relates a quantum system in $d-1$ (spatial) dimensions at $T=0$
and the corresponding
classic system in $d$ dimensions at finite temperature \cite{sachdev}. 
It generates the discretisation of the imaginary time-integral in momentum
space, i.e., summation on Matsubara frequencies \cite{negele,kapusta}.

To complete the finite-temperature RG description
one has to relate the temperature parameter to a momentum 
scale. In the perturbative RG, the standard choice is
$\mu = 2\pi T$ where $\mu$ is the RG scale. 
In the Wilsonian RG one has to take the limit $k\to 0$, and thus in
\cite{thermal_rg_phi4,thermal_rg_phi4_pressure,thermal_rg_phi4_volume}
the temperature 
is linked to the ultraviolet (UV)
initial value $\Lambda$ of the running momentum cutoff
by putting
$T = \tau \Lambda$.  
More precisely, the couplings are defined at an arbitrary but fixed
intermediate scale 
$k_\star = 2\pi T = 2\pi \tau \Lambda$ (so $T = \tau \Lambda$) and the
zero-temperature flow 
equation is integrated from $k_\star$ up 
to the UV cutoff $\Lambda$. Then starting from these bare parameters
one can follow the RG flow 
down from $k=\Lambda$ to $k=0$ 
with the temperature $T$ turned on. The physical 
quantities are then obtained at $k=0$. However, this choice has a 
consequence that 
the explicit $k$-dependence cannot be removed from the dimensionless
RG flow equations. 

Finite temperature formalism requires an introduction of a "finite volume" for 
the (imaginary) time integral. 
However, if one follows the Wilsonian approach, this "finite volume" 
has to be rescaled in every blocking step, which means it has to be linked
to the running momentum cutoff $k$. Since in the  RG approach the
infrared (IR) limit $k \to 0$ has to be taken, this
"finite volume" cannot be considered as the (inverse) temperature,
but it can rather be used as a running 
momentum cutoff: $T\equiv k_T$. Thus, we suggest and explore the following
identification
\begin{equation}
  T \equiv k_T = \tau k \, .
  \label{sugg}
\end{equation}  
In this case, the temperature of the expanding Universe is related to the
dimensionless quantity 
$\tau$ (and not to $k_T$) where $k_T$ is the running cutoff for thermal
fluctuations and $k$ is for the 
quantum ones.
In the Wilsonian approach fluctuations are taken into account by
the successive elimination (integration) of degrees of freedom above
these running cutoffs 
which are chosen to be different in order to
highlight the different contributions of thermal and quantum 
fluctuations. For $k_T \sim k$ (i.e.,~for $\tau  \sim 1$) one
cannot distinguish between thermal and 
quantum fluctuations.

In this work we
discuss the possibility of a thermal RG method with the special choice
(\ref{sugg}) $T = \tau k$, where we will argue that $\tau$ plays
the role of the temperature of the cosmic plasma. We 
show that in this case the dimensionless RG flow equations have no explicit 
$k$-dependence. In addition, we show how this modified thermal RG can 
be used to handle UV divergences of the running of the cosmological constant,
and to "solve the triviality", i.e., to generate a thermal phase transition
in terms of $\tau$ for the 
$\phi^4$ theory in $d=4$ Euclidean dimensions.

It is important to note that all computations of the present paper are performed 
in the Euclidean spacetime. In order to connect the results with cosmology and particle 
physics, one could speculate that there is the need for justification of the use of Euclidean 
metric. Indeed, there is an increasing interest in the literature to discuss whether an analytic 
continuation to Minkowski spacetime is possible in the framework of the Wilsonian RG 
method, see e.g., \cite{minkowski_frg}. However, let us note that lattice calculations are 
also performed in the Euclidean spacetime with huge amount of phenomenological 
applications, which serve as a strong support for the use of Euclidean instead of the 
Minkowski spacetime.

%===============
\section{Thermal RG equation}
%===============

The RG concept plays an important role in the description of physical systems across 
different scales. The Wilsonian RG method has been constructed to perform the 
renormalization non-perturbatively \cite{wilson}. In general, RG allows to treat the effects 
of fluctuations via the successive elimination of the degrees of freedom which lie above a 
running momentum cutoff $k$ and generates the momentum-shell functional RG flow equation,
i.e., the Wetterich equation \cite{eea_rg}. At zero-temperature it is formulated in Euclidean 
spacetime and stands for the running effective action $\Gamma_k[\phi]$ with its Hessian 
$\Gamma^{(2)}_k[\phi]$ where $k$ is the running momentum cutoff (i.e., the RG scale).
The Wetterich equation \cite{eea_rg} contains, an appropriately chosen regulator function
$R_{k}$, which is introduced to decouple slow modes with low momenta while leaving high 
momentum modes unaffected. The regulator ensures that the effective action captures all 
relevant fluctuations.

In the finite-temperature formalism, 
momentum integrals of the zero-temperature RG equation \cite{eea_rg} are 
modified by using Matsubara frequency summation 
$$\int d^dp \to T \sum_{\omega_n} \int d^{d-1}p$$ which is a summation over 
discrete imaginary frequencies  -- for bosonic degrees of 
freedom: $\omega_n = 2 \pi n T$. Typically, one assumes that the 
regulator function $R_k(p)$ of the zero-temperature RG method is independent of 
the Matsubara frequency, but otherwise the same as for the zero-temperature case,
where a usual choice is the Litim regulator $R_k(p) = (k^2 - p^2) \Theta(k^2 - p^2)$ 
\cite{Litim2000}. In the local potential approximation (LPA) when one does not consider 
any wave function renormalization and the couplings of the scaling potential $V_k(\phi)$ 
carry the RG scaling, the RG equation at finite temperature $T$ with the 
(frequency-independent) Litim regulator is written as \cite{thermal_rg_phi4}:
\beq
\label{thermal_opt}
k\partial_k V_k(\phi) = \frac{2 \alpha_{d-1}}{d-1} \, k^{d-1} 
T \sum_{n=-\infty}^{\infty} \frac{k^2}{k^2+ \omega_n^2 +\partial^2_{\phi} V_k(\phi)} \, ,
\eeq
where $k$ is the RG scale, $\alpha_d = \Omega_d/(2(2\pi)^d)$, and 
$\Omega_d = 2 \pi^{d/2}/\Gamma(d/2)$ is the $d$-dimensional solid angle. 
The summation can be performed \cite{thermal_rg_phi4}, which results in 
\begin{eqnarray}
\label{thermal_opt_rg}
k\partial_k V_k(\phi) = \frac{2 \alpha_{d-1}}{d-1} \, k^{d+1} 
\frac{\coth{\left(\frac{\sqrt{k^2+\partial^2_{\phi} V_k(\phi)}}{2T}\right)}}{2\sqrt{k^2+\partial^2_{\phi} V_k(\phi)}} \,,
\end{eqnarray}
where we used the identity $\coth(x/2) = \frac{\exp(x)+1}{\exp(x)-1}$.
Eq.~\eq{thermal_opt_rg} has been derived in \cite{thermal_rg_phi4},
where the authors performed a detailed analysis
of the scalar polynomial field theory by the
thermal functional RG method
which was also
discussed in
\cite{thermal_rg_phi4_pressure} with the inclusion of pressure and in
\cite{thermal_rg_phi4_volume} 
by taking into account volume fluctuations too.

If one relates the temperature to the UV cutoff $T = \tau \Lambda$ the 
dimensionless flow equations obtained from \eq{thermal_opt_rg} have 
an explicit $k$-dependence which is not the case for the zero-temperature 
RG equation \cite{eea_rg}. 
Using the identification (\ref{sugg})
we suggest the following thermal 
RG equation:
\beq
\label{thermal_opt_rg_k}
k\partial_k V_k(\phi) = \frac{2 \alpha_{d-1}}{d-1} \, k^{d+1} 
\frac{\coth{\left(\frac{\sqrt{k^2+\partial^2_{\phi} V_k(\phi)}}{2\tau k}\right)}}{2\sqrt{k^2+\partial^2_{\phi} V_k(\phi)}} \,.
\eeq
The thermal RG \eq{thermal_opt_rg_k} has a singularity structure identical
to its zero-temperature counterpart and determined by $k^2+\partial^2_{\phi} V_k = 0$. 
In \app{app1} we show that the dimensionless RG flow equations derived from 
\eq{thermal_opt_rg_k} have no explicit $k$-dependence. This is not the case if $T$ is fixed, 
i.e., not $k$-dependent. This implies that the choice (\ref{sugg}) can lead to non-trivial fixed points, 
unlike the choice of a non-running temperature. This observation is one of the main results of the
paper.

These results
can be used, among other possible applications,
to study the critical behavior:
{\it (i)} around the Wilson-Fisher 
fixed point of the polynomial quantum field theory in lower dimensions;
and {\it (ii)} around the Coleman fixed point of the periodic quantum field theory in $d=2$ dimensions. 
Our goal here is to consider the
$\phi^4$ quantum field theory in $d=4$ dimensions.

%===============
\section{Inflationary Cosmology}
%===============

The origin and the precise mechanism of cosmic inflation is one of the
most pressing questions 
in modern cosmology \cite{guth,sato,density-fluct,slow-roll_1,slow-roll_2},
for reviews 
see \cite{inflation_reviews_1,inflation_reviews_2}. In its simplest form,
a scalar field, called the 
inflation, is assumed to roll down slowly from a potential hill towards
its minimum which explains 
the cosmic microwave background radiation (CMBR) anisotropy and
the seeds of the large-scale 
structure. Particle physics can provide us with candidates for the inflaton
field, however, the complete 
understanding of particle physics origin remains
open problem.

The observation of the Higgs boson renewed research activity where the inflaton is associated
with the SM Higgs field \cite{higgs_inf_1,higgs_inf_2,higgs_inf_3,higgs_inf_4,higgs_inf_5} introduces 
a large non-minimal coupling $\xi$ between the Higgs boson and the Ricci curvature scalar 
\cite{higgs_inf_6,higgs_inf_7,higgs_inf_8,higgs_inf_9}. In case 
of non-minimal coupling, the Einstein frame where slow-roll study is performed and the Jordan-frame 
where the RG flow is considered are related to each other via a non-linear transformation. RG 
transformations generate an additional, $R^2$ term. However,
we do not consider here quantum effects 
for gravity, only for the scalar matter field, and thus we choose
\begin{equation}
\label{EH_and_matter}
S = \int d^4x \sqrt{-g} \left[\frac{m^2_p + \xi \phi^2}{2} R 
+ \frac{1}{2} g^{\mu\nu} \partial_\mu \phi \, \partial_\nu \phi -V(\phi)
\right] \nonumber
\end{equation}
where $m_p = 1/\sqrt{8\pi G}$ is the Planck mass, $\xi$ is the non-minimal coupling and $\phi$ 
is the scalar field where the field independent term is the cosmological constant:
\beq
\label{cosmo_const}
V(\phi = 0) \equiv m_p^2 \Lambda_{\mr{cosmo}},
\eeq
where $\Lambda_{\mr{cosmo}}$ denotes the cosmological constant and not
the UV momentum cutoff. We 
assume an expanding homogeneous and isotropic Universe (with a flat curvature), and with the 
Friedmann--Lema\^{\i}tre--Robertson--Walker metric, $\sqrt{-g}=\sqrt{-\det(g_{\mu\nu})}=a^3$ 
where the time-dependence of the scalar factor $a(t)$ is given by the Friedmann equation 
coupled to the equation of motion of the scalar field. This scenario is the
an "economical" one for the theoretical framework for slow-roll inflation
where the usual form for the Higgs-like 
potential is
\beq
\label{higgs}
V_{{\rm Higgs}}(\phi) = \frac{g_4}{4} \left(\phi^2 - v^2 \right)^2
\eeq
which is the SM symmetry breaking potential where $v$ is the vacuum expectation 
value (VeV) and $g_4$ is the quartic coupling
(dimensionless in $d=4$ 
dimensions). The slow-roll study of \eq{higgs} has to confronted
to observations, e.g., to 
most recent Planck data \cite{planck} on CMBR anisotropy. If one assumes a non-minimal 
coupling to gravity, i.e., $\xi \neq 0$, the potential has to be written in the Einstein frame to 
perform the slow-roll study. This transformation is non-linear and have a simplified form for 
large non-minimal coupling ($\xi \gg 1$): $\varphi \approx  m_p \sqrt{\frac{3}{2}} \ln(F)$ 
and $U(\varphi) \equiv  m_p^2 V(\phi)/F^2(\phi)$ where $F(\phi) = 1+ \xi \phi^2/m_p^2$. 
The Higgs-like potential \eq{higgs} reads in the
Einstein frame:
\begin{eqnarray}
\label{einstein_frame}
U_{\rm Higgs}(\varphi) = \frac{m_p^4 g_4}{4 \xi^2} 
\left[1 - \exp\left( -\sqrt{\frac23} \dfrac{\varphi}{m_p} \right) \right]^2,
\end{eqnarray}
where the VeV is neglected because it is assumed to be small compared to the field  
\cite{higg_inf_rubio}. The validity of the large $\xi$ approximation is
characterised by a critical value $\varphi_c$. Below this critical scale, Higgs inflation 
coincides, with the SM minimally coupled to gravity  \cite{higg_inf_rubio}. The large 
difference between the electroweak and the transition scales allows us to neglect the 
VeV \cite{higg_inf_rubio}, however it plays an important role 
in the electroweak phase transition.

%===============
\section{Thermal RG and Triviality in Cosmology}
%===============
We apply the thermal RG equation \eq{thermal_opt_rg_k} to the Higgs-like potential 
\eq{higgs} in $d=4$ dimensions by using 
\beq
V_k(\phi) = \sum_{n=0}^{\rm NCUT} \frac{g_{2n,k}}{(2n)!} \phi^{2n} 
\hskip 0.0cm \to \hskip 0.0cm
\tilde V_k(\tilde \phi) = \sum_{n=0}^{\rm NCUT} \frac{\tilde g_{2n,k}}{(2n)!} {\tilde \phi}^{2n}\, , 
\nonumber
\eeq
where dimensionless quantities denoted by the tilde superscript. Thermal RG 
flow equations for the first three (dimensionful) couplings are given in \app{app1}
where we show that the dimensionless RG flow equations have no explicit 
k-dependence and discuss also the subtraction method \cite{rg_cosmo_const} 
for the field-independent coupling $g_{0,k}$ which is related to the cosmological 
constant.

\begin{figure}[t!] 
\begin{center} 
\epsfig{file=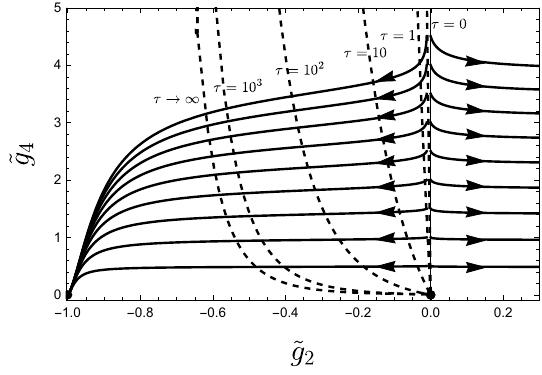,width=8.2cm}
\caption{\label{fig1} 
Thermal RG flow of the $\phi^4$  model in $d=4$ dimensions. Dashed
lines show how the separatrix changes with $\tau$. The separatrix 
belongs to $\tau=0$ (most right dashed line) almost overlaps with the vertical axis 
(thin solid line).
} 
\end{center}
\end{figure}

\begin{figure}[t!] 
\begin{center} 
\epsfig{file=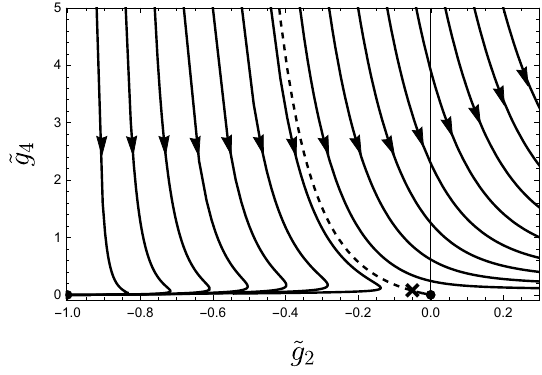,width=8.6cm}
\caption{\label{fig2} 
Thermal RG flow of the $\phi^4$  model in $d=4$ dimensions with 
$\tau = 10^2$. Black cross denotes an initial condition which lies on the 
separatrix with $\tau=10^2$. For $\tau < 10^2$ the RG trajectory from that 
starting point runs into the broken (low-temperature) phase and for $\tau > 10^2$ 
the RG trajectory ends up in the symmetric (high-temperature) phase.}
\end{center}
\end{figure}

It is well-known that the Wilson-Fisher fixed point of the $\phi^4$ model coincides 
with the Gaussian one in $d=4$ dimensions \cite{vbf_mati}. Although the model has two phases, 
RG trajectories bifurcate from the vertical line at the vanishing mass (when $\tilde g_{2,k} = 0$), 
see solid lines on \fig{fig1}. 
%
% Fig 1
%
The classical analysis gives exactly the same result: If the squared mass 
is negative the potential has a double-well structure which signals the broken
phase where the $Z_2$ reflection symmetry is broken spontaneously. If the 
squared mass is positive the potential has a single minimum which is the
symmetric phase. The numerical solution of the (dimensionless) thermal RG 
flow equations (see \app{app1}) modifies the flow diagram. With 
non-vanishing value for $\tau$,
the RG trajectory which separates the phases is no longer a vertical line, see
dashed lines on \fig{fig1}. For increasing values of $\tau$, the broken phase 
is decreased and for $\tau \to \infty$ it is not possible to find any starting point 
in the vicinity of the Gaussian fixed point from which an RG trajectory can run 
into the broken phase. Since $\tau$ measures how thermal fluctuations are
important compared to quantum fluctuations, one can say that by changing
$\tau$ a thermal phase transition occurs. One can always determine a critical 
value $\tau_c$ for any starting point close to the Gaussian fixed point, see the 
black cross on \fig{fig2}. If $\tau < \tau_c$ the RG 
trajectory from that starting point runs into the broken (low-temperature) 
phase and for $\tau > \tau_c$ the RG trajectory ends up in the symmetric 
(high-temperature) phase. Thus, the RG flow diagram is no longer "trivial" and 
a thermal phase transition is observed in terms of $\tau$. 
%
% Fig 2
%
The temperature of the Universe at the scale of cosmic inflation is many 
magnitudes higher than the temperature taken at the electroweak scale. The 
VeV is zero above and non-vanishing below the electroweak transition temperature. 
Although the SM symmetric phase requires the use of an SU(2) doublet Higgs field, 
in a simplified scenario one can use the thermal phase transition of the 4-dimensional 
$\phi^4$ model to switch between the broken and the symmetric phases. In other 
words, the zero or non-zero value of the VeV must distinguish between the low and 
the high temperature phases. Below the electroweak phase transition temperature, 
it is valid to use the SM symmetry breaking potential \eq{higgs} and consequently 
the 4-dimensional $\phi^4$ model.
Time and the thermal evolution of the Universe are linked to each other, so,
the couplings of the model are assumed to depend on the temperature which is 
encoded in their dependence on the parameter $\tau$. So, for an arbitrary but 
fixed starting point taken from the vicinity of the Gaussian fixed point, see again 
the black cross on \fig{fig2}, one can determine a critical value $\tau_c$ which can 
be associated to the critical temperature of the electroweak phase transition. The 
temperature at the scale of inflation is many magnitudes higher, $\tau \gg \tau_c$, 
thus the RG trajectory from that starting point runs into the symmetric 
(high-temperature) phase and consequently a vanishing VeV must be chosen 
in Eq.~\eq{higgs} which supports the use of Eq.~\eq{einstein_frame} in the 
slow-roll study of Higgs inflation.

%===============
\section{Conclusions}
%===============

A modification of the usual finite-temperature renormalization (RG) approach 
is proposed by relating the temperature parameter to the running RG scale:
$T \equiv k_T = \tau k$. This choice is convenient since with it
the dimensionless RG equation has no explicit $k$-dependence, see \app{app1}, 
and this is crucial to have fixed points. We applied this thermal RG to handle UV 
divergences in the RG running of the cosmological constant, see \app{app1}, and 
to solve the triviality $\phi^4$ theory in $d=4$. These results suggest to associate 
the parameter $\tau$ with the temperature of the expanding Universe, so 
$g_2 = g_2(\tau)$ and $g_4 = g_4(\tau)$ are assumed. The quantum effects 
are taken into account by the thermal RG running of the couplings at a given 
temperature, i.e., one has to take the limit $k\to 0$ by moving along 
the thermal RG trajectory with fixed value of $\tau$ with given initial 
conditions. Then $\tau$ can be decreased and the thermal 
RG procedure repeated, producing the quantum effective action at 
a given temperature. Time and thermal evolutions of the Universe 
are linked to each other, so the change in $\tau$ connects the scales of 
the cosmic inflation and the electroweak phase transition and thermal 
phase transition of the $\phi^4$ theory validates the vanishing VeV in 
Eq.~\eq{higgs} when the slow-roll study of Higgs inflation is performed
by Eq.~\eq{einstein_frame}. 

We observe that since perturbative 
and non-perturbative RG equations are related to each other 
\cite{scheme1,scheme2,scheme3,frg_prg}, one expects that our modified 
thermal RG method with $T \equiv k_T = \tau k$ finds application in particle 
cosmology both in perturbative and non-perturbative approaches. 
However, let us note that it is a valid question to ask whether the Wilsonian RG 
running scale parameter $k$ can or cannot be identified with physical properties 
of the system. In order to compute observables using the effective action one 
has to take the physical limit $k \to 0$ which is not yet possible in reasonable 
approximations.  Although, there are RG studies of cosmological problems where 
the Wilsonian flow with the scale "k" is used as an approximation to the physical flow 
with respect to the physical momentum "p", in this letter we relay on the effective 
action taken in the physical limit, i.e., $k \to 0$ form where physical quantities can 
be extracted. In other words, we do not consider the k-flow as an approximation 
to catch the main physics properties of the system at a given range of energy.

Finally, let us mention that an interesting direction of research would 
be to discuss $T\to \infty$ considering quantum gravity effects that should become 
relevant at the Planck scale, in order to understand how does the inclusion of gravity 
affects the proposed scheme. It would be also useful to apply the new thermal RG 
approach proposed here with the introduction of the dimensionless temperature 
$\tau$ to investigate in this context the QPT-CPT phase diagram of quantum field 
theories \cite{qpt_cpt_thermal_rg}, such as the $\phi^4$ and the sine-Gordon models 
in lower dimensions. The so called QPT-CPT diagram represents graphically the 
interplay between the classical phase transitions (CPT) and the quantum phase 
transitions (QPT) which has also been investigated in connection to the 
Naturalness/Hierarchy problem \cite{BrBrCoDa,BrBrCo,qpt_higgs1,qpt_higgs2}. 
For example, in \cite{qpt_higgs1,qpt_higgs2} its was shown that the hierarchy problem 
as well as the metastability of the electroweak vacuum can be understood as the Higgs 
potential being near-critical, i.e., close to a QPT.

%===============
\section*{Acknowledgements} 
%===============
Discussions with Z. Tr\'ocs\'anyi, T. Kov\'acs, N. Defenu and G. Fej\H os
and support for the CNR/MTA Italy-Hungary 2023-2025 Joint Project 
"Effects of strong correlations in interacting many-body systems and 
quantum circuits" are gratefully acknowledged.

\appendix

%------------------------------------------------------------------------------------ 
\section{Thermal RG flow equations of the $\phi^4$ model in $d=4$ Euclidean dimensions.}
\label{app1}
%------------------------------------------------------------------------------------
By using $T \equiv k_T = \tau k$ the following thermal RG equation 
can be written, see also \eq{thermal_opt_rg_k}
\beq
\label{app_thermal_opt_rg_k}
k\partial_k V_k(\phi) = \frac{2 \alpha_{d-1}}{d-1} \, k^{d+1} 
\frac{\coth{\left(\frac{\sqrt{k^2+\partial^2_{\phi} V_k(\phi)}}{2\tau k}\right)}}{2\sqrt{k^2+\partial^2_{\phi} V_k(\phi)}} \,.
\eeq
It can be rewritten for the dimensionless potential $\tilde V_k(\tilde \phi)$ with 
dimensionless field variable $\tilde \phi$ which are related to their dimensionful
counterparts as $\phi = k^{(d-2)/2} \tilde \phi$, $V_k = k^{d} \tilde V_k$ with
$\partial^2_{\phi} V_k(\phi) = k^2  \partial^2_{\tilde \phi} \tilde V_k(\tilde \phi)$.
The resulting thermal RG equation for dimensionless quantities reads as,
\begin{align}
\label{dimles_thermal_opt_rg_k}
\left(d - \frac{(d-2 )}{2} \tilde \phi \partial_{\tilde \phi} +  k\partial_k \right) & \tilde V_k(\tilde \phi) =
\\ \nonumber 
\frac{2 \alpha_{d-1}}{d-1} \, &
\frac{\coth{\left(\frac{\sqrt{1+ \partial^2_{\tilde \phi} \tilde V_k(\tilde \phi)}}{2\tau}\right)}}{2\sqrt{1+\partial^2_{\tilde \phi} \tilde V_k(\tilde \phi)}} \,,
\end{align}
which has no explicit k-dependence. However, if one takes the usual 
choice and relates the temperature parameter to a fixed momentum scale,
i.e., $T = \tau \Lambda$ one finds a dimensionless thermal RG equation 
with explicit k-dependence,
\begin{align}
\label{old_dimles_thermal_opt_rg_k}
\left(d - \frac{(d-2 )}{2} \tilde \phi \partial_{\tilde \phi} +  k\partial_k \right)& \tilde V_k(\tilde \phi) = 
\\ \nonumber
\frac{2 \alpha_{d-1}}{d-1} \, &
\frac{\coth{\left(\frac{k}{\Lambda} \frac{\sqrt{1+ \partial^2_{\tilde \phi} \tilde V_k(\tilde \phi)}}{2\tau}\right)}}{2\sqrt{1+\partial^2_{\tilde \phi} \tilde V_k(\tilde \phi)}} \,,
\end{align}
which makes no room for non-trivial fixed point solutions, i.e., by setting 
$k\partial_k \tilde V_k \equiv 0$, the remaining algebraic equation can have
only k-dependent solution. 

Eq.~\eq{app_thermal_opt_rg_k}
can be applied to a Higgs-like potential, 
in $d=4$ dimensions by using a general scale-dependent polynomial potential:
\beq
V_k(\phi) = \sum_{n=0}^{\rm NCUT} \frac{g_{2n,k}}{(2n)!} \phi^{2n} 
\hskip 0.0cm \to \hskip 0.0cm
\tilde V_k(\tilde \phi) = \sum_{n=0}^{\rm NCUT} \frac{\tilde g_{2n,k}}{(2n)!} {\tilde \phi}^{2n}\, , 
\nonumber
\eeq
where dimensionless quantities denoted by the tilde superscript. Thermal RG 
flow equations for the first three (dimensionful) couplings, i.e., for NCUT=2 are 
the following in $d=4$ dimensions:
\begin{align}
\label{g0_flow}
k \partial_k g_{0,k} = & \frac{1}{6 \pi^2} k^5    
\frac{\coth{\left(\frac{\sqrt{k^2+ g_{2,k}}}{2\tau k}\right)}}{2\sqrt{k^2+g_{2,k}}}, \\
\label{g2_flow}
k \partial_k g_{2,k} = & \frac{1}{6 \pi^2} k^5  
\left[
	-\frac{g_{4,k} \coth \left(\frac{\sqrt{k^2+g_{2,k}}}{2 \tau k }\right)}
		{4 \left(k^2+g_{2,k}\right)^{3/2}} 
	\right. \\ & \left. 
	-\frac{g_{4,k} \; \text{csch}^2\left(\frac{\sqrt{k^2+g_{2,k}}}{2 \tau k}\right)}
		{8 \tau k \left(k^2+g_{2,k}\right)} 
	\nonumber
\right], \\
\label{g4_flow}
k \partial_k g_{4,k} = & \frac{1}{6 \pi^2} k^5
\left[
	\frac{9 g_{4,k}^2 \coth \left(\frac{\sqrt{k^2+g_{2,k}}}{2 \tau k }\right)}
		{8 (k^2+g_{2,k})^{5/2}}
	\right. \\ & \left. 
	+
	\frac{9 g_{4,k}^2 \text{csch}^2\left(\frac{\sqrt{k^2+g_{2,k}}}{2 \tau k}\right)}
		{16 \tau k (k^2+g_{2,k})^2}
	\right. \nonumber \\ & \left. 
	+
	\frac{3 g_{4,k}^2 \coth \left(\frac{\sqrt{k^2+g_{2,k}}}{2 \tau k }\right)
		\text{csch}^2\left(\frac{\sqrt{k^2+g_{2,k}}}{2 \tau k}\right)}
		{16 \tau^2 k^2 (k^2+g_{2,k})^{3/2}}
	\nonumber
\right]
\end{align}
where $g_0$ is the field independent term. For dimensions $d=1$ this 
is associated with the ground-state energy, for $d=4$ it plays the role of 
the cosmological constant.

To perform the so called subtraction method \cite{rg_cosmo_const} for the finite 
temperature RG approach one has to consider the RG flow equation of the 
field-independent term in the UV limit ($k^2 \gg g_{2,k}$), i.e., the Taylor expansion 
of \eq{g0_flow} with respect to $g_{2,k}$ around zero:
\beq
k \partial_k g_{0,k} \approx \frac{k^4}{12\pi^2} \coth{\frac{1}{2\tau}} 
+ \frac{k^2}{24 \pi^2} \frac{1+\tau \sinh{\frac{1}{\tau}}}{\tau-\tau \cosh{\frac{1}{\tau}}} g_{2,k} + ... \, ,
\nonumber  
\eeq
where the first (second) term has a $k^4$ 
($k^2$) divergence, so they have to be subtracted in order to restore the 
Gaussian fixed point for the dimensionless couplings. The correct form of 
the thermal RG flow equation for $g_{0,k}$ is
\begin{align}
\label{g0_flow_sub} 
k \partial_k g_{0,k} = & \frac{1}{6 \pi^2} k^5 
\frac{\coth{\left(\frac{\sqrt{k^2+ g_{2,k}}}{2\tau k}\right)}}{2\sqrt{k^2+g_{2,k}}} 
- \frac{k^4}{12\pi^2} \coth{\frac{1}{2\tau}} 
\nonumber \\ &
- \frac{k^2}{24 \pi^2} \frac{1+\tau \sinh{\frac{1}{\tau}}}{\tau-\tau \cosh{\frac{1}{\tau}}} g_{2,k} \,,
\end{align}
where the subtracted terms do not violate the IR behaviour since they vanish 
for $k\to 0$. The RG flow for the cosmological constant can be derived from 
\eq{g0_flow_sub} based on the relation 
\beq
\label{cosmo_const}
V(\phi = 0) \equiv m_p^2 \Lambda_{\mr{cosmo}},
\eeq
where $\Lambda_{\mr{cosmo}}$ denotes the cosmological constant and not
the UV momentum cutoff, which gives
\begin{equation}
\label{flow_cosmo}
k\partial_k \lambda_k = 8\pi \left( g_k k \partial_k \tilde g_{0,k} + \tilde g_{0,k} k\partial_k g_k\right)
\end{equation}
with $\lambda_k = \Lambda_{\mr{cosmo},k} k^{-2}$, $g_k = G k^2$ is the dimensionful 
Newton constant, and $G$ is scale-dependent due to the absence of 
quantum gravity, so $k\partial_k g_k = 2 g_k$. By using dimensionless 
couplings, $g_{0,k} = \tilde g_{0,k} k^4$ and $g_{2,k} = \tilde g_{2,k} k^2$ 
the flow equation for the field independent term is substituted into 
\eq{flow_cosmo} and the thermal RG flow equation for the cosmological 
constant is obtained.

As a final step, let us rewrite the flow equations for dimensionless 
couplings $g_{0,k} = \tilde g_{0,k} k^4$, $g_{2,k} = \tilde g_{2,k} k^2$ and 
$g_{4,k} = \tilde g_{4,k}$ by using the subtracted form \eq{g0_flow_sub}. 
In this case one finds,
\begin{align}
(4 +  k \partial_k) \tilde g_{0,k} = & \frac{1}{6 \pi^2} 
\frac{\coth{\left(\frac{\sqrt{1 + \tilde g_{2,k}}}{2\tau}\right)}}{2\sqrt{1+\tilde g_{2,k}}} 
- \frac{1}{12\pi^2} \coth{\frac{1}{2\tau}} 
\nonumber \\ &
- \frac{1}{24 \pi^2} \frac{1+\tau \sinh{\frac{1}{\tau}}}{\tau-\tau \cosh{\frac{1}{\tau}}} \tilde g_{2,k}, 
\\
(2 +  k \partial_k) \tilde g_{2,k} = & \frac{1}{6 \pi^2}  \left[
	-\frac{\tilde g_{4,k} \coth \left(\frac{\sqrt{1+\tilde g_{2,k}}}{2 \tau}\right)}
		{4 \left(1+\tilde g_{2,k}\right)^{3/2}} 
	\right. \\ & \left. 
	-\frac{\tilde g_{4,k} \; \text{csch}^2\left(\frac{\sqrt{1+\tilde g_{2,k}}}{2 \tau}\right)}
		{8 \tau \left(1+\tilde g_{2,k}\right)} 
	\nonumber
\right], \\
k \partial_k \tilde g_{4,k} = & \frac{1}{6 \pi^2} 
\left[
	\frac{9 \tilde g_{4,k}^2 \coth \left(\frac{\sqrt{1+\tilde g_{2,k}}}{2 \tau}\right)}
		{8 (1+\tilde g_{2,k})^{5/2}}
	\right. \\ & \left. 
	+
	\frac{9 \tilde g_{4,k}^2 \text{csch}^2\left(\frac{\sqrt{1+\tilde g_{2,k}}}{2 \tau}\right)}
		{16 \tau (1+\tilde g_{2,k})^2}
	\right. \nonumber \\ & \left. 
	+
	\frac{3 \tilde g_{4,k}^2 \coth \left(\frac{\sqrt{1+\tilde g_{2,k}}}{2 \tau}\right)
		\text{csch}^2\left(\frac{\sqrt{1+\tilde g_{2,k}}}{2 \tau}\right)}
		{16 \tau^2 (1+\tilde g_{2,k})^{3/2}}
	\nonumber
\right].
\end{align}
Thus, we demonstrated that by using our proposal $T \equiv k_T = \tau k$ the 
dimensionless thermal RG flow equations have no explicit RG scale-dependence,
so one can find non-trivial fixed points. This is not the case if $T$  is fixed, not $k$-dependent. 
Therefore, with $T = \tau k$ the usual RG flow diagram method can 
be used to study the critical behaviour.

We finally observe that perturbative and non-perturbative RG 
equations are related to each other \cite{scheme1,scheme2,scheme3,frg_prg}, thus, one 
expects that our modified thermal RG method with $T \equiv k_T = \tau k$ finds 
application in particle cosmology both in perturbative and non-perturbative approaches.


\begin{thebibliography}{97}

\bibitem{wilson}
K. G. Wilson, % {\it Renormalization group and critical phenomena. I. Renormalization group and the Kadanoff scaling picture}, 
Phys. Rev. B {\bf 4}, 3174 (1971); 
K. G. Wilson, %{\it Renormalization group and critical phenomena. II. Phase-space cell analysis of critical behavior}, 
Phys. Rev. B {\bf 4}, 3184 (1971).

\bibitem{reuter}
M. Reuter, %{\it Nonperturbative evolution equation for quantum gravity}, 
Phys. Rev. D. {\bf 57}, 971 (1998). 

\bibitem{reuter_saueressig}
M. Reuter, F. Saueressig, {\it Quantum Gravity and the Functional Renormalization Group: The Road towards Asymptotic Safety} 
(Cambridge, %Monographs on Mathematical Physics,
Cambridge University Press, 2019).

\bibitem{AS_grav_1}
A. Eichhorn, M. Schiffer, {\it Invited chapter for the "Handbook of Quantum Gravity"}, arXiv:2212.07456 [hep-th].

\bibitem{AS_grav_2}
A. Platania, {\it Special issue "Coarse Graining in Quantum Gravity: Bridging the Gap between Microscopic Models and Spacetime-Physics"}, 
arXiv:2003.13656 [gr-qc].

\bibitem{bonanno_saueressig}
A. Bonanno, F. Saueressig, %{\it Asymptotically safe cosmology - a status report}, 
C. R. Phys. {\bf 18}, 254 (2017).

\bibitem{alessia}
A. Platania, %{\it From renormalization group flows to cosmology}, 
Front. Phys. {\bf 8}, 188 (2020).

\bibitem{rg_cosmo_12}
A. Babic, B. Guberina, R. Horvat, H. Stefancic, %{\it Renormalization-group running cosmologies: A scale-setting procedure}, 
Phys. Rev. D {\bf 71}, 124041 (2005).

\bibitem{rg_cosmo_17}
M. Hindmarsh, I. D. Saltas, %{\it $f(R)$ gravity from the renormalization group}, 
Phys. Rev. D {\bf 86}, 064029 (2012). 

\bibitem{rg_cosmo_18}
E. J. Copeland, C. Rahmede, I. D. Saltas, %{\it Asymptotically safe Starobinsky inflation}, 
Phys. Rev. D {\bf 91}, 103530 (2015).

\bibitem{cosmo_qeg}
G. Gubitosia, R. Ooijera, C. Ripkena, F. Saueressig, %{\it Consistent early and late time cosmology from the RG flow of gravity},
J. Cosmol. Astropart. Phys. {\bf 1805}, 004 (2018).

\bibitem{qeg8}
A. Platania, F. Saueressig, %{\it Functional Renormalization Group Flows on Friedman--Lemaitre--Robertson--Walker backgrounds},
Found. Phys. {\bf 48}, 1291 (2018).

\bibitem{negele}
J. W. Negele, H. Orland, {\it Quantum many-particle systems} (Reading, Perseus, 1998).

\bibitem{kapusta}
  J. I. Kapusta, {\it Finite-temperature field theory: principles and applications} (Cambridge, Cambridge University Press, 2023).

\bibitem{sachdev}
S. Sachdev, {\it Quantum phase transitions} (Cambridge, Cambridge University Press, 2011).
  
\bibitem{thermal_rg_phi4}
J.-P. Blaizot, A. Ipp, R. M\'endez-Galain, N. Wschebor, %{\it Perturbation theory and non-perturbative renormalization flow in scalar field theory at finite temperature},
Nucl. Phys. A {\bf 784}, 376 (2007). 

\bibitem{thermal_rg_phi4_pressure}
J.-P. Blaizot, A. Ipp, N. Wschebor, %{\it Calculation of the pressure of a hot scalar theory within the Non-Perturbative Renormalization Group}, 
Nucl. Phys. A {\bf 849} 165 (2011).

\bibitem{thermal_rg_phi4_volume}
L. Fister, J. M. Pawlowski, %{\it Functional renormalization group in a finite volume}
Phys. Rev. D {\bf 92}, 076009 (2015).

\bibitem{eea_rg}
C. Wetterich, %{\it Exact evolution equation for the effective potential}, 
Phys. Lett. B {\bf 301}, 90 (1993);  
T. R. Morris, %{\it The Exact renormalization group and approximate solutions}, 
Int. J. Mod. Phys. A {\bf 9}, 2411 (1994).

\bibitem{Litim2000}
D. F. Litim, %{\it Optimisation of the exact renormalisation group}, 
Phys. Lett. B {\bf 486}, 92 (2000).

\bibitem{guth}
A. H. Guth, %{\it The Inflationary Universe: A Possible Solution to the Horizon and Flatness Problems},
Phys. Rev. D {\bf 23}, 347 (1981).

\bibitem{sato}
K. Sato, %{\it First Order Phase Transition of a Vacuum and Expansion of the Universe}, 
Mon. Not. Roy. Astron. Soc. {\bf 195}, 467 (1981).

\bibitem{density-fluct}
A. A. Starobinsky, %{\it A new type of isotropic cosmological models without singularity}, 
Phys. Lett. B {\bf 91}, 99 (1980);
V. F. Mukhanov, G. V. Chibisov, %{\it Quantum fluctuations and a nonsingular universe}, 
JETP Lett. {\bf 33}, 532 (1981) [Pisma Zh. Eksp. Teor. Fiz. {\bf 33}, 549 (1981)].

\bibitem{slow-roll_1}
A. D. Linde, %{\it A New Inflationary Universe Scenario: A Possible Solution of the Horizon, Flatness, Homogeneity, Isotropy and Primordial Monopole Problems},
Phys. Lett. B {\bf 108}, 389 (1982).

\bibitem{slow-roll_2}
A. Albrecht, P. J. Steinhardt, %{\it Cosmology for Grand Unified Theories with Radiatively Induced Symmetry Breaking},
Phys. Rev. Lett. {\bf 48}, 1220 (1982).

\bibitem{inflation_reviews_1}
D. H. Lyth, A. Riotto, %{\it Particle physics models of inflation and the cosmological density perturbation}, 
Phys. Rep. {\bf 314} 1 (1999).

\bibitem{inflation_reviews_2}
D. Baumann, {\it Inflation}, in Theoretical Advanced Study Institute in Elementary Particle Physics: Physics of the Large and the Small, 523. (2011), arXiv:0907.5424 [hep-th];
D. Baumann, %{\it Primordial Cosmology}, 
{\it PoS} {\bf TASI2017}, 009 (2018), arXiv:1807.03098 [hep-th].

\bibitem{higgs_inf_1}
T. Futamase, K.-i. Maeda, %{\it Chaotic Inflationary Scenario in Models Having Nonminimal Coupling With Curvature}, 
Phys. Rev. D {\bf 39}, 309 (1989).

\bibitem{higgs_inf_2}
R. Fakir, W. G. Unruh, %{\it Improvement on cosmological chaotic inflation through nonminimal coupling}, 
Phys. Rev. D {\bf 41}, 1783 (1990).

\bibitem{higgs_inf_3}
J. L. Cervantes-Cota, H. Dehnen, %{\it Induced gravity inflation in the standard model of particle physics}, 
Nucl. Phys. B {\bf 442},391 (1995).

\bibitem{higgs_inf_4}
F. L. Bezrukov, M. Shaposhnikov, %{\it The Standard Model Higgs boson as the inflaton}, 
Phys. Lett. B {\bf 659}, 703 (2008).

\bibitem{higgs_inf_5}
F. Bezrukov, J. Rubio, M. Shaposhnikov, 
Phys. Rev. D {\bf 92} 083512 (2015).

%================

\bibitem{higgs_inf_6}
C. P. Burgess, H. M. Lee, M. Trott,  % Power-counting and the Validity of the Classical Approximation During Inflation,
JHEP {\bf 09}, 103 (2009).

\bibitem{higgs_inf_7}
J. L. F. Barbon, J. R. Espinosa, % On the Naturalness of Higgs Inflation,
Phys. Rev. D {\bf 79}, 081302 (2009).

\bibitem{higgs_inf_8}
C. P. Burgess, H. M. Lee, M. Trott,  % Comment on Higgs Inflation and Naturalness,
JHEP {\bf 07}, 007 (2010).

\bibitem{higgs_inf_9}
M. P. Hertzberg, % On Inflation with Non-minimal Coupling,
JHEP {\bf 11}, 023 (2010).

%================

\bibitem{higg_inf_rubio}
J. Rubio, %{\em Higgs Inflation}, 
Front. Astron. Space Sci. {\bf 5}, 50 (2019).

\bibitem{planck}
N. Aghanim et al. {\em et al.} [Planck collaboration], {\it Planck 2018 results.  VI. Cosmological parameters}, Astron. Astrophys. {\bf 641}, A6 (2020),
[Erratum: Astron. Astrophys. {\bf 652}, C4 (2021)].

\bibitem{rg_cosmo_const}
I. G. M\'ari\'an, U. D. Jentschura, N. Defenu, A. Trombettoni, I. N\'andori, %{\em Vacuum energy and renormalization of the field-independent term},
J. Cosmol. Astropart. Phys. {\bf 03}, 062 (2022).

\bibitem{vbf_mati}
P. Mati, % Vanishing beta function curves from the functional renormalization group
Phys. Rev. D {\bf 91}, 125038 (2015).

\bibitem{scheme1}
A. Codello, M. Demmel, O. Zanusso, Phys. Rev. D {\bf 90}, 027701 (2014).

\bibitem{scheme2}
C. Branchina, V. Branchina, F. Contino, N. Darvishi, Phys. Rev. D {\bf 106}, 065007 (2022).

\bibitem{scheme3}
C. Branchina, V. Branchina, F. Contino, Phys. Rev. D {\bf 107}, 096012 (2023).

\bibitem{frg_prg}
S. Hariharakrishnan, U. D. Jentschura, I. G. M\'ari\'an, K. Szab\'o, I. N\'andori,
J. Phys. G: Nucl. Part. Phys. {\bf 51}, 085005 (2024).

%================

\bibitem{minkowski_frg}
I. Steib, S. Nagy, J. Polonyi, Int. J. Mod. Phys. A {\bf 36}, 2150031 (2021); 
F. G\'eg\'eny, S. Nagy, Int. J. Mod. Phys. A {\bf 36}, 2250061 (2022).

\bibitem{qpt_cpt_thermal_rg}
I. G. M\'ari\'an, A. Trombettoni, I. N\'andori, arXiv: 2407.20704 [hep-th].

\bibitem{BrBrCoDa}
C. Branchina, V. Branchina, F. Contino, N. Darvishi, Phys. Rev. D {\bf 106}, 065007 (2022).

\bibitem{BrBrCo}
C. Branchina, V. Branchina, F. Contino, Phys. Rev. D {\bf 107}, 096012 (2023).

\bibitem{qpt_higgs1}
T. Steingasser, arXiv:2405.02415 [hep-ph]. %, DOI: https://doi.org/10.22323/1.463.0150

\bibitem{qpt_higgs2}
T. Steingasser, David I. Kaiser, Phys. Rev. D {\bf 108}, 095035 (2023).



\end{thebibliography}
\end{document}